\documentclass[prl,aps,reprint,superscriptaddress,nofootinbib]{revtex4-1}
\usepackage[utf8]{inputenc}
\usepackage[english]{babel}
\usepackage{amssymb}
\usepackage{amsmath}
\usepackage{amsfonts}
\usepackage{mathtools}
\usepackage{graphicx}
\usepackage{float}
\usepackage{dsfont} 
\usepackage{cancel}
\usepackage{enumitem}
\usepackage{appendix}
\usepackage{simplewick}
\usepackage{appendix}
\usepackage{bbm}
\usepackage{xcolor}
\usepackage{hyperref}

\newcommand{\bea}{\begin{eqnarray}}
\newcommand{\eea}{\end{eqnarray}}

\newcommand{\PP}{\mathcal{P}}
\newcommand{\cev}[1]{\reflectbox{\ensuremath{\vec{\reflectbox{\ensuremath{#1}}}}}}

\begin{document}

\title{Quantum Computation as Gravity}
\begin{flushright}
YITP-18-75
\end{flushright}
\author{Pawe{\l} Caputa}
\affiliation{Center for Gravitational Physics, Yukawa Institute for Theoretical Physics (YITP),\\Kyoto University, Kitashirakawa Oiwakecho, Sakyo-ku, Kyoto 606-8502, Japan.}

\author{Javier M. Magan}
\affiliation{Instituto Balseiro, Centro Atomico Bariloche S. C. de Bariloche, Rio Negro, R8402AGP, Argentina}

\begin{abstract}
We formulate Nielsen's geometric approach to complexity in the context of two-dimensional conformal field theories, where series of conformal transformations are interpreted as "unitary circuits". We show that the complexity functional can be written as the Polyakov action of two-dimensional gravity or, equivalently, as the geometric action on the coadjoint orbits of the Virasoro group. This way, we argue that gravity sets the rules for optimal quantum computation in conformal field theories. 

\end{abstract}

\maketitle
%%%%%%%%%%%%%%%%%
{\bf I. Introduction and Summary.} Information, be it classical or quantum, is intimately related to geometry. This is manifest in the field of computational complexity, which, as shown by Nielsen et all. \cite{Nielsen1,Nielsen2,Nielsen3}, is naturally framed in the language of differential geometry. The main idea is that computational costs can be estimated by distances in the manifold of unitaries. Therefore, minimizing costs amounts to finding minimal geodesics, and optimal algorithms are given by a free fall through such `complexity geometry'. Being fundamentally geometric, it is tempting to expect deep relations with General Relativity (GR).\\
On the other hand, the AdS/CFT correspondence \cite{Maldacena:1997re} assures us that complexity in conformal field theories (CFT), once properly defined, should be encoded geometrically in Anti-de Sitter spacetimes and may, in fact, be crucial to fully understand holography \cite{SusskindQC}. Up to date, we have some holographic proposals for what "complexity" may be and they usually declare some gravitational notion, such as spacetime volumes or gravity actions, to be related to the state complexity, see \cite{SusskindQC,SStanford,slaw,Miyaji2015,Brown1,Brown2,Aaronson,Chapman:2016hwi,Alishahiha:2015rta,Lehner:2016vdi}. However, it is far from obvious how to define complexity in CFTs. Recent proposals and advances in this direction include \cite{Miyaji2015,Caputa2017yrh,Czech,Myers,Chapman,Yang,Yang2,Khan,Abt2017pmf,Hashimoto2017fga,Molina-Vilaplana2018sfn, Abt2017pmf,Hackl,Flory2018akz,Magan2018}). \\
With the motivation of deepening these connections, we resort to the underlying symmetries in order to build bridge(s) between the two fields and develop Nielsen's geometric approach within 2d CFT. Building upon the ideas in \cite{Magan2018}, we focus on a subset of quantum symmetry gates that, in 2d CFTs, are fully governed by the Virasoro group. The setup universality allows us to express the complexity action only in terms of the CFT central charge $c$. Moreover, we show that the ambiguity of choosing a metric on Virasoro circuits is a $1/c$ effect, and at large-c the complexity action is given by Polyakov's two-dimensional gravity \cite{Polyakov:1987zb}. This means that gravity governs the rules of optimal computation within this class of symmetry gates in 2d CFTs. On a similar footing, since Polyakov and Liouville actions are directly related, our results neatly connect to the recent proposal for path integral complexity \cite{Caputa2017yrh}.\\
More importantly, it is also a well-known fact that Polyakov action is equivalent to the coadjoint orbit action of the Virasoro group \cite{Alekseev:1988ce}. This connection between quantum complexity and coadjoint orbit actions is one of the main results of our work. It allows concrete generalizations of our story to general coherence groups in arbitrary quantum field theories, as described at the end of the letter.\\
We believe that the present approach reintroduces in a conceptually simple way important aspects of CFT's which, albeit known and scattered through a relatively older (string theory) literature, can set the quantum complexity discussion on firmer ground. Crucially, all the needed building blocks (the generators of the Virasoro algebra) can be defined on both sides of the duality, so holographic complexity proposals should rest on geometric relations between bulk objects.\\
Last but not least, our considerations certainly open a new interesting path in the field of quantum complexity in many-body systems.\\
%%%%%%%%%%%%%%%%%%%%%%%%%
{\bf II. Nielsen Complexity and Symmetry.} We start by reviewing the geometric approach to complexity based on \cite{Nielsen1,Nielsen2,Nielsen3} and  \cite{Magan2018}.
The goal is to estimate computational costs of quantum circuits $U(\tau)$, which in their most generic form are defined as
\begin{equation}
U(\tau)=\cev{\PP}\exp\left[-i\int^\tau_0H(\tau')d\tau'\right],\label{Ut}\;,
\end{equation}
driving us from a given reference state at $\tau=0$ to a particular target state at $\tau$. The Hermitian operators $H(\tau)$ depend on the allowed operations that we have at our disposal (quantum gates) and $\cev{\PP}$ denotes gate ordering (earlier $\tau$ first).\\
Nielsen's attractive approach to estimate the associate costs is via metrics on the (group) manifold of unitaries. In this context, complexity becomes the length of the shortest geodesic between $U(0)=1$ and $U(\tau)$.\\
More precisely, we can decompose any circuit into infinitesimal gates $U_{H(\tau)}\equiv e^{-iH(\tau)d\tau}$, with $H(\tau)$ living in the tangent space at point $\tau$. In differential geometry metrics are norms on the tangent space, so the most general Nielsen's cost is given by 
\begin{equation}\label{Hcost}
\mathcal{C}(\tau)=\int\limits^{\tau} \,\mathcal{F}(H(\tau'))\,d\tau'\;,
\end{equation}
where $\mathcal{F}$ is some norm on the tangent space. In general, it is useful to think about such cost actions as describing a particle on a group manifold (e.g. sigma model type action given by length or energy), with associated equations of motion given by the geodesic equation.\\
Before defining $\mathcal{F}$, let us expand on the protagonist of this whole story, which is $H(\tau)$. We will call $H(\tau)$ `the instantaneous gate', since it takes us from $U(\tau)$ to $U(\tau+d\tau)$:
\begin{eqnarray}\label{Hs}
&U(\tau+d\tau)=e^{-i H(\tau)d\tau}U(\tau).& 
\end{eqnarray}
Such instantaneous gates are the `velocities' in the group manifold. In general, their actual computation involves infinite sums of nested commutators (see e.g. \cite{kirillovbook}), and it is impossible to write a closed form suitable for computing complexity. To make progress, in \cite{Magan2018} one of us proposed to study submanifolds associated to symmetry groups $G$. In this scenario, gates are unitary representations $U_{g}$ of group elements $g$, and continuous protocols $U_{g(\tau)}$ are defined by paths $g(\tau)$ in the group. The instantaneous gate equation~(\ref{Hs}) has an homologus group equation:
\begin{equation}
g(\tau+d\tau)=e^{Q (\tau)d\tau}\cdot g(\tau)\;,\label{SymGa}
\end{equation}
where $g\cdot g'$ denotes the group product. The instantaneous gate is thus a Lie algebra element $-iH(\tau)=Q(\tau)$ (in the appropriate representation). For the mathematically oriented readers, $Q (\tau)$ is the adjoint transformation of the Maurer-Cartan (MC) form, known for many Lie groups, including the Virasoro group.\\
Next, we need to define the cost function $\mathcal{F}$ which is non-unique. Nielsen's proposal  \cite{Nielsen1}, in the context of spin systems, concerns two canonical choices, namely the one- and two- norms $\mathcal{F}_1$ and $\mathcal{F}_2$ , which can be defined as the first and second moments of the infinitesimal gate in the maximally mixed state. This definition seems problematic in the case of continuous systems but, as discussed in \cite{Magan2018}, for symmetry gates this can be circumvented by replacing the "maximally mixed state" with the actual density matrix
\begin{equation}
\rho(\tau)\equiv U(\tau)\rho_0U^\dagger(\tau),
\end{equation}
for some reference state $\rho_0$ that can be pure or mixed. Then, we define the norms as expectation values of the instantaneous symmetry gate(s) $Q(\tau)$ as
\begin{eqnarray}\label{cost}
&&\mathcal{F}_{1}(\tau)\equiv|\textrm{Tr}(\rho (\tau)Q (\tau))|=|\textrm{Tr}(\rho_0\tilde{Q} (\tau))|,\nonumber\\
&&\mathcal{F}_{2}(\tau)\equiv\sqrt{-\textrm{Tr}(\rho (\tau)Q (\tau)^2)}=\sqrt{-\textrm{Tr}(\rho_0 \tilde{Q} (\tau)^2)},
\end{eqnarray}
where both $Q (\tau)$ and $\tilde{Q}(\tau)\equiv U^\dagger(\tau)Q(\tau)U(\tau)$ are assumed to be given in the appropriate representation $U_{g}$. It is important to stress that, up to an overall constant, these costs are fixed by symmetry considerations. Moreover, not only being well defined from spin systems to interacting quantum field theories, these costs are now physical quantities, the actual moments of the instantaneous rate of change of the state.\\ 
Finally, the complexity action (the length) is computed by~(\ref{Hcost}) with either $\mathcal{F}_{1}$ or $\mathcal{F}_{2}$. As we will see below, this construction naturally fits into the framework of CFTs. Moreover, in the large central charge limit (holographic CFTs), both metrics become equivalent  $\mathcal{F}_2\simeq \mathcal{F}_1 (1+\mathcal{O}(1/c))$.\\
%%%%%%%%%%%%%%%%%%%%%%%%%
{\bf III. Circuit Complexity in 2d CFTs.} Now we consider the geometric approach to complexity in arguably the simplest setup relevant for holography, namely 2d CFTs (see \cite{DiF} for references). We consider computational tasks in the (symmetry) unitary manifold that is at the core of every CFT, namely that of the Virasoro group.\\
The Virasoro group is the central extension of the group of diffeomorphisms of the circle $ \text{Diff}^+ (S^1)$ preserving orientation. Group elements are maps $f(\sigma)$ that are $2\pi$-periodic $f(\sigma+2\pi)=f(\sigma)+2\pi$ and invertible $f'(\sigma)>0$, as well as their group product is given by the composition
\begin{equation}
f\cdot g= f\circ g.
\end{equation}
In 2d CFT we have two copies of the Virasoro group that appear as the transformations of the cylinder light-cone coordinates. These diffeomorphisms are represented in the Hilbert space by unitary operators $U_f$, yielding~e.g.~the well known transformation of the stress tensor with Schwarzian derivative (see e.g. \cite{DiF}). Since our arguments are purely group theoretic, we first focus on a single copy.\\
To be precise, Virasoro elements are pairs $(f,\alpha)$ where $f\in \text{Diff}^+ (S^1)$ and $\alpha\in \mathbb{R}$. The product is the composition of functions and addition of numbers with the Bott cocycle (see \cite{Witten:1987ty} and e.g. \cite{Oblak:2016eij} for pedagogical review). In the complexity context, we will not care about these numerical  phases or, in other words, we associate cost zero to the identity gate.\\
Following the previous sections, we consider a unitary CFT circuit build from the Virasoro symmetry gates $U_{f}$ (unitary representations of the Virasoro group). Our Virasoro circuit $U(\tau)$ defines a path $f(\tau,\sigma)$ in the group manifold, taking us from a reference state $\left|\psi_R\right>$ for $f(0,\sigma)=\sigma$ to a target state $\left|\psi_T\right>=U_{f(\tau,\sigma)}\left|\psi_R\right>$ for $f(\tau,\sigma)$  (to be clear: label $\tau$ means that for each $\tau$ we have one diffeomorphism $f_\tau(\sigma)\equiv f(\tau,\sigma)$).\\
The circuit is defined as
\begin{equation}
U(\tau)=\cev{\PP}\exp\left[\int^\tau_0Q(\tau')d\tau'\right],\label{UtVir}
\end{equation}
where the instantaneous gate, that belongs to the Virasoro Lie algebra, can be expressed in terms of the stress tensor (see supplementary material for our conventions)
\begin{equation}\label{gateV}
Q(\tau)\equiv\int^{2\pi}_0\frac{d\sigma}{2\pi}\epsilon(\tau,\sigma)  T(\sigma)=\sum_{n\in \mathbb{Z}}\epsilon_n(\tau) \left(L_{-n}-\frac{c}{24}\delta_{n,0}\right),
\end{equation}
where we Fourier expanded  $\epsilon(\tau,\sigma)$ as well as the stress tensor on the cylinder in terms of the Virasoro algebra generators:
\begin{equation}\label{Virasoro}
[L_m,L_n]=(m-n)L_{m+n}+\frac{c}{12}m(m^2-1)\delta_{m+n,0}.
\end{equation}
The unitarity of our circuit is guaranteed by the condition $\epsilon^*_n(\tau)=-\epsilon_{-n}(\tau)$. However, as described in the supplementary material, this condition can be extended so that our circuits give the most general protocols in the CFT vacuum sector.\\
To compute the cost function, we need to relate $\epsilon(\tau,\sigma)$ (velocities) to the path $f(\tau,\sigma)$ itself. This is where the symmetry becomes crucial. Namely, since the group product is given by composition, by definition of the infinitesimal gate~(\ref{SymGa}) we must have $\epsilon(\tau,f(\tau,\sigma))=\partial_\tau f(\tau,\sigma)$ or equivalently
\begin{equation}
\epsilon(\tau,\sigma)=\partial_\tau f(\tau,F(\tau,\sigma))=-\frac{\partial_\tau F(\tau,\sigma)}{\partial_\sigma F(\tau,\sigma)},\label{EPSF}
\end{equation}
where we introduced the inverse function $F(\tau,f(\tau,\sigma))=\sigma$. This explicit form of velocities allows us to express complexity (for any definition of the cost) as a functional of the path $f(\tau,\sigma)$. From now on we omit arguments of $f$'s and denote $\partial_\tau f=\dot{f}$ and $\partial_\sigma f=f'$ (and similarly for $F$).\\
Now we need to specify the reference state $\rho_0$. We take it to be a pure eigenstate of the CFT hamiltonian created by a primary operator with (chiral) dimension $h$: $\rho_0=|h\rangle\langle h|$. Since a basis in CFT can be obtained by acting with strings of $L_n$'s on $|h\rangle$ we find this to be the most natural choice (see also next section where this state specifies on which coadjoint orbit we are). Employing the Virasoro algebra~(\ref{Virasoro}) and using the transformation law of the stress tensor, we compute the one-norm
\begin{eqnarray}
\mathcal{F}_1(\tau)&=&\frac{c}{24\pi}\int^{2\pi}_0d\sigma\frac{\dot{f}}{f'}\left(2a^2+\{f,\sigma\}\right)
\end{eqnarray}
where $2a^2=\frac{1}{2}\left(1-\frac{24h}{c}\right)$. Similarly we can evaluate the two-norm (see supplementary material for the exact result).  It turns out to be
\begin{eqnarray}
\mathcal{F}_2(\tau)&=&\mathcal{F}_1(\tau)(1+\mathcal{O}(1/c)),
\end{eqnarray}
as could have been anticipated due to large-c factorization. The two choices become equivalent in the large-c limit, giving hope that some of the non-universal ambiguities of circuit complexity are "washed out" in holographic CFTs. Note also that the one norm choice can be interpreted as the norm of the Berry connection for the Virasoro group.\\
This way, our large-c complexity functional for paths $f(\tau,\sigma)$ between the identity and a given final transformation $f(\tau,\sigma)=f(\sigma)$ becomes  
\begin{equation}\label{CFTc1}
\mathcal{C}[f](\tau)=\frac{c}{24\pi}\int^\tau_0d\tau'\int^{2\pi}_0 d\sigma \frac{\dot{f}}{f'}\left(2a^2+\{f,\sigma\}\right).
\end{equation}
A simple but meaningful optimal protocol appears for $SL(2,R)$ paths with vanishing Schwarzian derivative. Then for the class of paths where $\dot{f}/f'=c_1$ is constant, we arrive to the linear complexity growth 
\begin{equation}\label{LinGr}
\mathcal{C}[f](\tau)=c_1 E_h\, \tau,
\end{equation}
where $E_h=|h-\frac{c}{24}|$ is the expectation value of the energy-momentum tensor. More complicated examples will be explored in \cite{CM}.\\
Finally, performing analogous steps for the second copy of the Virasoro group, the full CFT complexity action reads
\begin{equation}\label{CFTc2}
\mathcal{C}^{CFT}[f,\bar{f}](\tau)=\mathcal{C}[f](\tau)+\mathcal{C}[\bar{f}](\tau).
\end{equation}
where $\bar{f}$ is the analogous (independent) conformal path for the left Virasoro group. Above we have used $Q=Q_{L}+Q_{R}$. Had we defined a cost mediated by a negative sign $Q=Q_{L}-Q_{R}$, we would have obtained $\mathcal{C}[f]-\mathcal{C}[\bar{f}]$. At the level of the classical solutions there is no difference between both choices.\\
We remark again that the cost is written in terms of Lie algebra elements. In AdS/CFT, these are boundary operators, with known dual fields. Therefore, our costs are built from well defined quantities at both sides of the AdS/CFT duality. In fact we will now discuss more direct connection between our complexity and gravity.\\
%%%%%%%%%%
{\bf IV. 2d Gravity.} The connection of the 2d-CFT complexity functional~(\ref{CFTc1}) to gravity appears via the Polyakov's action of (induced) gravity in two dimensions \cite{Polyakov:1987zb}:
\begin{equation}
S_P[g]=\frac{c}{24\pi}\int d^2x\sqrt{g}\left(-\frac{1}{4}R\frac{1}{\Box}R+\Lambda\right),\label{PolA}
\end{equation}
where $R$ is the Ricci scalar curvature, $\Lambda$ the cosmological constant and $\Box$ the Laplace-Beltrami operator (its inverse $1/\Box$ acting on $R$ is just a formal expression implying that $R\equiv\Box(...)$). Polyakov action is intimately related to Virasoro symmetry, and it plays the role of the generating functional of the stress tensor correlators
\begin{equation}\label{equivalence}
e^{-S_P[\mu]}\equiv\langle e^{-\frac{1}{2\pi}\int d\tau d\sigma\, \mu\, \mathbf{T}}\rangle,
\end{equation}
where $S_P[\mu]$ is the Polyakov action computed in metric (Polyakov gauge) \cite{Polyakov:1987zb}
\begin{equation}\label{Gauges}
ds^2=d\tau\left(d\tilde{\sigma}+\mu(\tau,\tilde{\sigma})\, d\tau\right)=G'(\tau,\sigma)d\tau d\sigma,
\end{equation}
where $\mu=\dot{g}/g'$ and $g(\tau,G(\tau,\sigma))=\sigma$.\\
This way, we have e.g. the one-point function
\begin{equation}
\frac{\delta}{\delta \mu}S_P[\mu]=\frac{1}{2\pi}\langle\mathbf{T} \rangle=\frac{c}{24\pi}\{g(\tau,\sigma),\sigma\}.
\end{equation}
In our story, the essential step is the large-c limit after which the expectation value of the exponent becomes the exponent of the expectation value of our infinitesimal gate. Hence, our complexity becomes the Polyakov action in a convenient gauge (choice of the metric). In the quantum information language, induced 2d gravity governs the complexity of Virasoro circuits! This is the main result of our work and a completely new light on this well established core of 2d CFTs.\\
Let us elaborate more on this important connection and bring a new light that allows for generalizations towards complexity in quantum field theories.\\
 The Polyakov action evaluated on metric \eqref{Gauges} can be compactly written as
\begin{equation}
S_P[G]=\frac{c}{24\pi}\int d\tau\int d\sigma \frac{\dot{G}}{2G'}\left(\frac{G'''}{G'}-2\frac{G''^2}{G'^2}\right).\label{PolGg}
\end{equation}
and, as proven in \cite{Haba:1989in,Aldrovandi:1996sa}, it leads to the same e.o.m and expectation value of $\mathbf{T}$ as the one with full Schwarzian derivative  $S_P[\mu]\sim \int \mu \{g,\sigma\}$.  Were we on the plane, this would be our large-c complexity functional. As shown in \cite{Alekseev:1990mp,Barnich:2017jgw}, on the cylinder we just need to substitute $G=\exp(\sqrt{2}aF)$ and using $f(\tau,F(\tau,\sigma))=\sigma$ one arrives to \eqref{CFTc1}. This makes the precise connection with the functions $F$ and $f$ (and parameter $a$) used in our complexity functional.\\
Let us point that for two Virasoro copies we get two Polyakov/Complexity actions. It is well known \cite{Henneaux,Henneaux:1999ib} how they can be combined into a single, non-chiral, Liouville theory. This sheds new light on the Path Integral Complexity proposal \cite{Caputa2017yrh}. It is an important future problem to compare the details of the two approaches \cite{CM}   (see also \cite{Milsted:2018vop,Milsted:2018yur,Molina-Vilaplana:2015rra} for related constructions).\\
Finally, as described a long time ago in \cite{Alekseev:1988ce}, Polyakov action  \eqref{PolGg} is in fact the Kirillov geometric action \cite{kirillovbook} on the coadjoint orbits of the Virasoro group. In the next section we describe this relation in detail, since, as will be discussed in the last section, it gives a hint and a definite route for constructing complexity (based on "generalized symmetry gates") in arbitrary quantum field theories.\\
%%%%%%%%%%%%%%%%%%%%%%%%%
{\bf V. Geometric Actions.} \label{sec:GeomActions} Geometric actions, developed by Kirillov \cite{kirillovbook}, arise naturally in the coadjoint orbit method in representation theory. They play a crucial role in the context of geometric quantization. The framework starts with a group $G$ and a path in the group $g(\tau)$. Associated to such path we have the instantaneous gate  defined as
\begin{equation}\label{Q}
Q (\tau)= \frac{d}{ds}(g (s)\cdot g^{-1}(\tau) )\vert_{s=\tau}\;,
\end{equation}
and the Maurer-Cartan form:
\begin{equation}
\tilde{Q} (\tau)=\frac{d}{ds}( g^{-1}(\tau) \cdot g (s))\vert_{s=\tau}\;.
\end{equation} 
They both belong to the Lie algebra $\mathfrak{g}$ of $G$. We have also the dual space $\mathfrak{g}^{*}$, the space of linear functionals from $\mathfrak{g}$ to the reals, denoted by $\langle v, Q\rangle \in \mathds{R}$, where $v\in \mathfrak{g}^{*}$ and $Q\in \mathfrak{g}$. 

As it is well known, there is a group action on the Lie algebra, the adjoint representation
\begin{equation}\label{adjoint}
Ad_{g}(Q)=\frac{d}{d\tau}(g\cdot e^{\tau Q} \cdot g^{-1})\vert_{\tau =0}\;.
\end{equation}
It induces the coadoint representation on the dual space $\mathfrak{g}^{*}$, implicitly defined by:
\begin{equation}\label{coadjoint}
\langle Ad^{*}_{g}(v), Q\rangle \equiv \langle v, Ad_{g^{-1}}(Q)\rangle \;.
\end{equation}
Now, given an element $v\in \mathfrak{g}^{*}$ and a group path $g(\tau)$, the geometric action is defined as:
\begin{equation}
I_{\textrm{Geometric}}=p \int d\tau \,\langle Ad^{*}_{g(\tau)} v , Q(\tau)\rangle =p \int d\tau \,\langle v , \tilde{Q}(\tau)\rangle \;.
\end{equation}
The intuition is that $v$ defines a state, as in~(\ref{cost}). The action is just the `average' of the instantaneous velocity $Q$ in the evolved state $Ad^{*}_{g(\tau)} v$. The connection between this action and the one-norm complexity definition~(\ref{cost}) should now be transparent.

A deeper understanding of the previous construction goes as follows. The coadjoint representation defines orbits in $\mathfrak{g}^{*}$. Each orbit, represented by some $v\in \mathfrak{g}^{*}$, contains all elements that can be reached from $v$ by a coadjoint transformation. If the stabilizer of $v$ is $H\subset G$, the fundamental observation, see \cite{kirillovbook} and \cite{Oblak:2016eij}, is that each orbit $G/H$ is a symplectic manifold, where an invariant non-degenerate symplectic form can be defined, the Kirillov-Kostant form:
\begin{equation}
\Omega=\frac{1}{2}\langle v, d \tilde{Q}\rangle\;.
\end{equation}
This form is closed and so locally $\Omega=d\alpha$, with $\alpha=\langle v, \tilde{Q}\rangle$. Therefore, the coadjoint orbit $G/H$ defines a phase space, with natural action:
\begin{equation}
I_G[g,v]=p\int \alpha=p \int d\tau \, \langle v , \tilde{Q}(\tau)\rangle\;.
\end{equation}
This method was applied to the Virasoro group in \cite{Alekseev:1988ce,Witten:1987ty}. The geometric action is:
\begin{equation}\label{geof}
I_{\textrm{Virasoro}}=\frac{c}{24\pi}\int d\tau\,d\sigma \frac{\dot{f}}{f'}\left(\frac{12}{c}b(f)+\frac{1}{2} \left( \frac{f''}{f'}\right)' \right).
\end{equation}
The $b(f)$ is an element of the dual space of the Virasoro Lie algebra and labels the specific coadjoint orbit. We can identify it with the stress tensor expectation value $b(f)=\textrm{Tr} (\rho_0T(f))$ and in our case of the primary state $b(f)=h-\frac{c}{24}=-\frac{c}{6}a^2$ (in general the dependance on $\sigma$ just means that $b(f)$ transforms as the expectation value of $T(\sigma)$). The second contribution is the first term of the Schwarzian derivative and, as functional for $f$, this action is equivalent to \eqref{CFTc1}.
Aspects of these actions were used recently in connection to SYK \cite{Stanford:2017thb,Mandal:2017thl,Mertens:2018fds} and Virasoro Berry phases \cite{Oblak:2017}.\\
%%%%%%%%%%%%%%%%%%%%%%%%%
{\bf VI. Generalizations and Discussion.} The most promising generalization relies on the observed connection between complexity and geometric actions. Such geometric actions appear ubiquitously through the notion of generalized coherent states, see \cite{coherent1,coherent2,yaffe,usdf}. Basically, for any quantum theory there exists a `coherence group' $G$. This group provides a classical phase space for the theory, by generating a continuous and normalizable Hilbert space basis, the coherent states $\vert g_{v}\rangle =U(g)\vert v\rangle$. In the semiclassical limit, the dynamic is determined by an action:
\begin{equation}\label{sol}
S=S_{\textrm{geometric}}+S_{\textrm{H}}=-\int_{\gamma}\alpha -\int d\tau \, H(\gamma (\tau))\;,
\end{equation}
where $\gamma (\tau)$ is a path on the coherence group and $H(\gamma (\tau))$ is the Hamiltonian (a functional on phase space). In this context, natural gates are infinitesimal elements of the coherence group, states are elements of the dual space and protocols are paths $\gamma (\tau)\in G$ defining paths in phase space by means of the coadjoint transformation. Using the same notion of cost~(\ref{cost}), in the semiclassical limit the complexity is again given by the geometric action:
\begin{equation}\label{geog}
\mathcal{C}=\int_{\gamma}\alpha
\end{equation}
By comparing the semiclassical solution~(\ref{sol}) and the complexity or geometric action~(\ref{geog}), the analysis suggests that for theories controlled by a Hamiltonian constraint $H(\gamma (\tau))=0$ the complexity action is exactly the classical action. This might connect nicely to gravitational theories.\\
Besides, in the context of AdS/CFT, this coherent state approach can be defined on both sides of the duality. This is because the instantaneous gates are smeared versions of local boundary operators, with known gravity duals, and associated Kirillov-Kostant symplectic forms are functionals of them (see e.g. \cite{Belin:2018fxe}). An important subset of the coherence group is $e^{i\int f(x)g(x)}$, with $g$ the metric, dual to  $e^{i\int f(x)T(x)}$, with $T$ the stress tensor. In two dimensions this generates the conformal group and has been considered above. We leave the higher dimensional case and the inclusion of matter sources for future research.\\
Related generalizations within 2d CFTs arise when allowing instantaneous gates to include other primaries or symmetry currents (like Kac-Moody or higher spin W), or when including supersymmetry. Indeed, geometric actions for the Kac-Moody symmetry are well known and the approach of Alexeev and Shatashvili was also generalized to $W_3$ (see e.g. \cite{Marshakov:1989ca,Ooguri:1991by}).\\
To finish, let us stress that the present approach provides a definite starting point to derive gravity from complexity. In our setup, the CFT complexity corresponds to the Virasoro geometric action for each left and right sectors. Ref \cite{Alekseev:1988ce,Forgacs:1989ac} showed how each chiral sector is equivalent to a $SL(2,\mathbf{R})$ WZW theory. In turn, these WZW theories are the reduction at the boundary of $AdS_3$ of the two Chern-Simons actions of 3d gravity \cite{WittenS}. \\
W can further elaborate on the 2d-3d gravity connection. Firstly, solutions of 3d gravity are fixed by solutions of the 2d Liouville equation. The later can be written in terms of our $\text{Diff}(S^1)$ diffeomorphisms. Indeed, from a non-chiral Liouville field, we can construct left and right stress tensors completely specifying a three-dimensional metric \cite{Banados:1998gg}. Secondly, the alluded Liouville action also computes the area of a two-dimensional surface, naturally associated with a slice of a dual bulk geometry (see \cite{Caputa2017yrh} for Euclidean slices). Even though it is by no means clear which slice this may be, we conjecture that it is always possible to find such a slice in the bulk.\\
%%%%%%%%%%%%%
%%%%%%%%%%%%%
{\bf Acknowledgements.} We wish to thank Hugo Camargo, Horacio Casini, Sumit Das, Jan de Boer, Nilay Kundu, Hugo Marrochio, Gautam Mandal, Rob Myers, Masamichi Miyaji, Tokiro Numasawa, Blagoje Oblak, Onkar Parrikar, Fernando Pastawski, Massimo Poratti, Guifre Vidal and especially Joan Simon and Tadashi Takayanagi for guidance and important suggestions. Our work is supported by the Simons Foundation through the ``It from Qubit'' collaboration and work of PC was also supported by the JSPS starting grant KAKENHI 17H06787. We thank the Galileo Galilei Institute for Theoretical Physics for the hospitality and the INFN for partial support during the completion of this work.

%%%%%%%%%%%%%%%%%%%%%%%%%%%%%%%%%%%
%%%%%%%%%%%%%%%%%%%%%%%%%%%%%%%%%%%

%(SUPPLEMENTARY MATERIAL)

\newpage

\appendix{}
%%%%%%%%%%%%%%%%%%%%%%%%%%%%%
\section{Details of our setup}\label{appx:Setup}
%%%%%%%%%%%%%%%%%%%%%%%%%%%%%
As we argued in the main text, in 2d CFTs, we can define quantum circuits built form the Virasoro symmetry gates as
\begin{equation}
U(\tau)=\cev{\PP}\exp\left[\int^\tau_0\left(Q(\tau')+\bar{Q}(\tau')\right)d\tau'\right],\label{UtCFT}
\end{equation}
with the instantaneous gate defined as
\begin{equation}
Q(\tau)\equiv\int^{2\pi}_0\frac{d\sigma}{2\pi}\epsilon(\tau,\sigma)  T(\sigma)=\sum_{n\in \mathbb{Z}}\epsilon_n(\tau) \left(L_{-n}-\frac{c}{24}\delta_{n,0}\right),
\end{equation}
where we have used the expansion on the cylinder of size $2\pi$
\begin{eqnarray}
T(\sigma)&=&\sum_{n\in \mathbb{Z}} \left(L_n-\frac{c}{24}\delta_{n,0}\right)e^{ -i n \sigma},\nonumber\\
\epsilon(\tau,\sigma)&=&\sum_{n\in \mathbb{Z}} \epsilon_n(\tau)e^{ -i n \sigma},
\end{eqnarray}
and the $\delta$ functions
\begin{eqnarray}
\delta_{n+m,0}&\equiv&\int^{2\pi}_0\frac{d\sigma}{2\pi}e^{i(n+m)\sigma},\nonumber\\
\delta(\sigma_1-\sigma_2)&\equiv&\frac{1}{2\pi}\sum_{n\in \mathbb{Z}} e^{in(\sigma_1-\sigma_2)}.
\end{eqnarray}
Analogously, we have the Virasoro generators and Fourier coefficients
\begin{eqnarray}
L_n&=&\int^{2\pi}_0\frac{d\sigma}{2\pi}T(\sigma)e^{in\sigma}+\frac{c}{24}\delta_{n,0},\nonumber\\
\epsilon_n(\tau)&=&\int^{2\pi}_0\frac{d\sigma}{2\pi}\epsilon(\tau,\sigma)e^{in\sigma}.
\end{eqnarray}
Generators $L_n$ ($\bar{L}_n$) satisfy the algebra (\ref{Virasoro}).\\ 
Next, to compute the cost functions, we use the transformed gate
\begin{equation}
\tilde{Q}(\tau)\equiv\int^{2\pi}_0\frac{d\sigma}{2\pi}\epsilon(\tau,\sigma)U^\dagger_f(\tau)T(\sigma)U_f(\tau).
\end{equation}
where
\begin{equation}
U^\dagger_f T(\sigma)U_f\to f'^{-2}(\tau,\tilde{\sigma})\left(T(f(\tau,\tilde{\sigma}))-\frac{c}{12}\{f(\tau,\tilde{\sigma}),\tilde{\sigma}\}\right),
\end{equation}
The important point is that for the Virasoro symmetry gates, we can use \eqref{SymGa} to fix the dependence of $\epsilon(\tau,\sigma)$ on the group elements $f(\tau,\sigma)$.  Namely we have from  \eqref{SymGa}
\begin{equation}
f(\tau+d\tau,\sigma)=(1+\epsilon(\tau,\sigma)d\tau)\circ f(\tau,\sigma),
\end{equation}
which implies
\begin{equation}
\partial_\tau f(\tau,\sigma)d\tau=\epsilon(\tau,\sigma)\circ f(\tau,\sigma)d\tau=\epsilon(\tau,f(\tau,\sigma))d\tau.
\end{equation}
This is precisely the relation used in \eqref{EPSF}.\\
Note that our symmetry circuit is very general and we can e.g. write the part in the exponent as (suppressing the $(\tau,\sigma)$ in $\epsilon$'s)
\begin{equation}
Q(\tau)+\bar{Q}(\tau)=\int^{2\pi}_0d\sigma\left[\frac{\epsilon+\bar{\epsilon}}{2}h(\sigma)+i\frac{\epsilon-\bar{\epsilon}}{2i}p(\sigma)\right],
\end{equation}
where $2\pi\, h(\sigma)=T(\sigma)+\bar{T}(\sigma)$ and $2\pi\,p(\sigma)=T(\sigma)-\bar{T}(\sigma)$, such that the CFT Hamiltonian and momentum are
\begin{eqnarray}
H&=&\int^{2\pi}_0d\sigma h(\sigma)=L_0+\bar{L}_0-\frac{c}{12},\nonumber\\
P&=&\int^{2\pi}_0d\sigma p(\sigma)=L_0-\bar{L}_0.
\end{eqnarray}
This way, by imposing certain additional constraints on $\epsilon$ (and $\bar{\epsilon}$) our circuits can be seen as e.g. Euclidean or Lorenzian Path Integrals. But we would like to stress that our framework is more general. Below, we perform the analysis for the single copy of the Virasoro circuit.
%%%%%%%%%%%%%%%%%%%%%%%%%%%%%
\section{Different choices of the complexity metric}\label{appx:metric}
%%%%%%%%%%%%%%%%%%%%%%%%%%%%%
Let us discuss some alternative possibilities for choosing the complexity metric from our unitary circuit of diffeomorphisms. Let us start with the most general approach. Namely, given a continuous unitary circuit $U(t)$ as in \eqref{Ut} we can formally write (using summation convention)
\begin{equation}
\partial_\tau U(\tau )U(\tau )^{-1}=-i\tilde{H}(\tau)\equiv \epsilon^n(\tau)M_n
\end{equation}
where $M_n$ form some algebra specified by allowed set of gates. In the case of Virasoro gates they are infinite number of the Virasoro generators $M_n=L_n$. The key problem when computing complexity is to find the velocities $\epsilon^n(\tau)$ as a function of the unitary path. But once this step is completed, we can define a generic metric by
\begin{equation}
ds^2=G_{mn}\epsilon^m\epsilon^nd\tau^2,
\end{equation}
where $G_{mn}$ is arbitrary and can include "penalty factors" that penalize certain gates relatively to others (for example local gates may be "cheaper" than non-local ones).\\
From this metric we can define the cost functional as the length
\begin{equation}
\mathcal{C}(\tau)=\int^\tau ds=\int^\tau\sqrt{G_{mn}\epsilon^m(\tau')\epsilon^n(\tau')}d\tau'.\label{ComAcAp}
\end{equation}
The minimal cost, i.e. complexity, would then be the on-shell value of this action on the solution of the geodesic equation 
\begin{equation}
\dot{\epsilon}^\alpha+\Gamma^\alpha_{mn} \epsilon^m\epsilon^n=0.
\end{equation}
This is the standard equation remembering that $\epsilon^n=\dot{x}^n$ for some $x$ as it should for velocity.\\
The same equation of motion can be obtained from the "energy" type action
\begin{equation}
S[\tau]=\frac{1}{2}\int^\tau G_{mn}\epsilon^n(\tau')\epsilon^m(\tau')d\tau'.
\end{equation}
More precisely, we can introduce a Lagrange multiplier $e$ into the action
\begin{equation}
\tilde{S}[\tau,e]=\frac{1}{2}\int^\tau \left(e^{-1}G_{mn}\epsilon^n(\tau')\epsilon^m(\tau')+e\right)d\tau',
\end{equation}
such that the minimum is $e=|\epsilon|=\sqrt{G_{mn}\epsilon^m\epsilon^n}$ at which action becomes the length \eqref{ComAcAp}. This action is reparametrization invariant $\tau\to \tilde{\tau}(\tau)$ with $e\to e/(\tilde{\tau}'(\tau))$, so we can set $e=1$ by a proper choice of time variable. The main point now is that we can use each of these actions as complexity functional and they lead to the same equations of motion (geodesic). Moreover they all agree on the minimum path, hence give the same value of "complexity", provided that we parametrize the path by its arc-length.\\

In the main part of this paper, we derived the velocities for the Virasoro circuits where
\begin{equation}
\partial_\tau U(\tau)U(\tau)^{-1}=\epsilon^n(\tau)L_n,
\end{equation}
and the velocities are
\begin{equation}
\epsilon^n(\tau)=-\int^{2\pi}_0\frac{d\sigma}{2\pi}\frac{\dot{F}(\tau,\sigma)}{F'(\tau,\sigma)}e^{in\sigma}.
\end{equation}
Then in the above language, our one-norm metric can be written as
\begin{equation}
G_{mn}=\textrm{Tr}(\,\rho (\tau)\,L^\dagger_{m})\textrm{Tr}(\,\rho (\tau)\,L_{n})\;,
\end{equation}
whereas our two-norm metric is just
\begin{equation}
G_{mn}=\textrm{Tr}(\,\rho (\tau)\,L_{m}^{\dagger}L_{n})\;,
\end{equation}
where again $\rho (\tau)$ would be the actual state of the system and in both formulas $L_n$'s are the generators on the cylinder so $L_0\equiv L_0-\frac{c}{24}$.

Another natural (and closely related) metric would be the Fubini-Study metric. If we have the following protocol
\begin{equation}
|\Psi(\tau)\rangle=U(\tau)|R\rangle
\end{equation}
satisfying
\begin{equation}
|\partial_\tau\Psi(\tau)\rangle\equiv\partial_\tau|\Psi(\tau)\rangle=-i\tilde{H}(\tau)|\Psi(\tau)\rangle.
\end{equation}
then we can compute our previous cost
\begin{equation}
\langle\partial_\tau \Psi(\tau)|\partial_\tau\Psi(\tau)\rangle=\langle \Psi(\tau)|\tilde{H}(\tau)^2|\Psi(\tau)\rangle,
\end{equation}
as well as the Berry connection
\begin{equation}
\langle\Psi(\tau)|\partial_\tau\Psi(\tau)\rangle=-i\langle \Psi(\tau)|\tilde{H}(\tau)|\Psi(\tau)\rangle.
\end{equation}
The Fubini-Study metric is then defined as
\begin{equation}
ds^2=(\langle\partial_\tau \Psi(\tau)|\partial_\tau\Psi(\tau)\rangle-|\langle\Psi(\tau)|\partial_\tau\Psi(\tau)\rangle|^2)d\tau^2.
\end{equation}
Notice that in our previous notation $Q(\tau)=-i\tilde{H}(\tau)$ and we can write
\begin{equation}
ds^2=-(\langle R|\tilde{Q}(\tau)^2|R\rangle-(\langle R|\tilde{Q}(\tau)|R\rangle)^2)d\tau^2,
\end{equation}
where $\tilde{Q}(\tau)$ is the transformed infinitesimal gate. For our Virasoro gates, and again using as a reference state a primary one $|R\rangle=|h\rangle$, we have the following result for the Berry connection
\begin{eqnarray}
\langle h|\tilde{Q}(\tau)|h\rangle&=&-\frac{c}{24\pi}\int^{2\pi}_0d\sigma\frac{\epsilon(\tau,\sigma)}{f'^2(\tau,\sigma)}\left(2a^2+\{f(\tau,\sigma),\sigma\}\right)\nonumber\\
\end{eqnarray}
where $2a^2=\frac{1}{2}\left(1-\frac{24h}{c}\right)=\frac{12}{c}b(f)$. To arrive to such expression we used
\begin{equation}
\langle h|T(f)|h\rangle=-\frac{c}{24}\left(1-\frac{24h}{c}\right)\equiv b(f).
\end{equation}
Similarly, the second moment of the instantaneous gate can be written as
\begin{eqnarray}
\langle h|\tilde{Q}(\tau)^2|h\rangle&=&\left(\frac{c}{24\pi}\right)^2\left[\left(\int^{2\pi}_0d\sigma\frac{\dot{f}}{f'}\left(2a^2+\{f,\sigma\}\right)\right)^2\right.\nonumber\\
&+&\left.\frac{12}{c}\sum_{n\in \mathbb{Z}}p^h_n\int^{2\pi}_0d\sigma_1d\sigma_2\epsilon_1\epsilon_2\left(\frac{\sigma_2}{\sigma_1}\right)^n\right]\nonumber\\
\end{eqnarray}
where $\epsilon_i=\dot{f}_i/f'_i$ and $f_i=f(\tau,\sigma_i)$. This expression is our exact two-norm cost which approaches the square of the Berry connection at large-c. On the other hand, Fubini-Study would give the second term, based on the connected 2-point correlator of the stress-tensor
\begin{equation}
ds_{\textrm{FS}}^2=\frac{c}{48\pi^{2}}\sum_{n\in \mathbb{Z}}p^h_n\int^{2\pi}_0d\sigma_1d\sigma_2\epsilon_1\epsilon_2\left(\frac{\sigma_2}{\sigma_1}\right)^n
\end{equation}
which is $\mathcal{O}(1/c)$ in relation to the square of the Berry connection (our complexity functional). This way the complexity funcitonal would be proportional to $\sqrt{c}$. Clearly, in this particular case, the Fubini-Study choice gives a much worse lower bound to the quantum complexity of the protocol.

Such a hierarchical separation between the cost we have used to derive gravity (the generic one proposed in \cite{Magan2018}) and the Fubini-Study choice is expected to appear quite generically. The reason is simple. First, notice that for any instantaneous gate $H$ we have
\begin{equation}
\langle H^{2}\rangle\geq \langle H^{2}\rangle -\langle H\rangle^{2}\;,
\end{equation}
so Fubiny-Study gives at best an equally good choice. But notice that, physically, what we are comparing is the second moment of an observable $\langle H^{2}\rangle$ with its mean squared deviation $\sigma_{H}^{2}$. Usual statistical arguments show that $\sigma_{H}^{2}$ is subleading in an appropriate thermodynamic expansion. For example, if we are to compute the cost of Hamiltonian evolution $U(t)=e^{-iH t}$ in a high energy state, we need to compare $\langle H^{2}\rangle \simeq E^{2}(1+\mathcal{O}(T/E)) $ with $\sigma_{H}^{2}\propto T E$, which show different scalings with the entropy of the system, with $\sigma_{H}^{2}$ subleading.\\
From a parallel perspective, our cost would give
\begin{equation}
C(t)=Et\;,
\end{equation}
which is the so-called Lloyd's bound, whereas Fubiny-Study would give:
\begin{equation}
C_{\textrm{FS}}(t)=\sigma_{E}t\;,
\end{equation}
which is known as the Margolus-Levitin bound. The arguments just expressed above are a direct generalization of the relation between these two bounds.
%%%%%%%%%%%%%%%%%%%%%%%%%%%%%
\section{Euler-Arnold and Virasoro}\label{appx:EAV}
%%%%%%%%%%%%%%%%%%%%%%%%%%%%%
Last but not least, we could also take a more abstract approach to Virasoro circuits using Euler-Arnold equations from (mathematical) classical mechanics that we now review (see e.g. [AK] that we follow closely). \\
The Virasoro algebra $vir=\text{Vect}(S^1)\oplus \mathbb{R}$ consists of pairs $\left(f(\sigma)\partial_\sigma,c\right)$ where the basis in the vector space is defined by $\{L_n=ie^{in\sigma}\partial_\sigma\}$ where $n\in \mathbb{Z}$. On this space, we can define an $L^2$-inner product as
\begin{equation}
\langle \left(f(\sigma)\partial_\sigma,c_1\right),\left(g(\sigma)\partial_\sigma,c_2\right)\rangle=\int_{S^1}f(\sigma)g(\sigma)d\sigma+c_1\cdot c_2.
\end{equation}
The dual Virasoro algebra $vir^*$ consists of pairs, $\left(u(\sigma)d\sigma^2,c\right)$, of quadratic differentials and real numbers. The pairing between the algebras is given by 
\begin{equation}
\langle \left(g(\sigma)d\sigma^{2},c_2\right ),\left(f(\sigma)\partial_\sigma,c_1\right)\rangle=\int_{S^1}f(\sigma)g(\sigma)d\sigma+c_1\cdot c_2.
\end{equation}
This inner product allows us to define a map between the Lie algebra and the dual space, the "Inertia operator":
\begin{equation}
A:\mathfrak{g}\to\mathfrak{g}^{*}:\,\,\, \left(f(\sigma)\partial_\sigma,c\right)\to \left(f(\sigma)d\sigma^{2},c\right)\;,
\end{equation}
together with its inverse $A^{-1}:\mathfrak{g}^{*}\to\mathfrak{g}$.\\
We can now define a right-invariant metric on the Virasoro group, namely the Hamiltonian 
\begin{equation}
H\left(f(\sigma)\partial_\sigma,c\right)=\frac{1}{2}\left(\int_{S^1}f^2(\sigma)d\sigma+c^2\right)
\end{equation}
and similarly on the $vir^*$
\begin{equation}\label{HKd}
H\left(g(\sigma)d\sigma^2,c\right)\equiv\frac{1}{2}\langle \left(g(\sigma)d\sigma^2,c\right),A^{-1}\left(g(\sigma)d\sigma^2,c\right)\rangle.
\end{equation}
The Euler-Arnold equation on $\mathfrak{g}^{*}$ is given by
\begin{eqnarray}
\frac{d}{dt}m(t)=ad^*_{A^{-1}m(t)}m(t),\\
\nonumber
\end{eqnarray}
where $m(t)=(u(t,\sigma)d\sigma^2,c(t))$ is a point in $\mathfrak{g}^{*}$. The coadjoint action (see [AK]) can be written as
\begin{equation}
ad^*_{(u(t,\sigma)\partial_\sigma,c)}(u(t,\sigma)d\sigma^2,c)\equiv -((3uu'+cu''')d\sigma^2,0),
\end{equation}
such that the Euler-Arnold equation becomes the famous integrable equation, the KdV equation
\begin{equation}
\partial_t u=-3uu'-cu''',\qquad \partial_t c=0.
\end{equation}
Moreover, without the central extension ($c=0$) the geodesic equation becomes the Burger's equation.\\
In physical applications of KdV $u(t,\sigma)$ is usually the velocity field and looking back into section V, we can indeed identify $u(\tau,\sigma)=\epsilon(\tau,\sigma)$ for our Virasoro circuits. The consequences of these metrics for complexity will be explored in our future works.

\end{document}